\documentclass[journal,onecolumn]{IEEEtran}  
\usepackage{amsmath,amsfonts,amssymb}
\usepackage{graphicx}
\usepackage{xstring}
\usepackage{graphics}
\usepackage{graphics,color}
\usepackage{subcaption} 
\usepackage{cite}
\usepackage{url}
\usepackage{epstopdf}
\usepackage{times}

\usepackage[T1]{fontenc}
\usepackage[latin1]{inputenc}
\usepackage{times}

\usepackage{url}
\usepackage{stackrel}
\usepackage{epsf}
\usepackage{amsthm}
\usepackage{amsfonts}
\usepackage{graphicx}
\usepackage{graphics,color}
\usepackage{multirow}
\usepackage[english]{babel}
\usepackage{amsmath,amssymb}
\usepackage{cite}
\usepackage{times}
\usepackage{tabularx}
\usepackage{tikz}
\usetikzlibrary{matrix}
\usepackage{cite}
\usepackage{ifthen}
\usepackage{times}
\usepackage{tabularx}
\usepackage{epstopdf}
\usepackage{bm}
\input{epsf.sty}

\newcommand{\hide}[1]{\ifthenelse{\boolean{false}}{#1}{}}

\newtheorem{theorem}{{\bf Theorem}}
\newtheorem{lemma}{{\bf Lemma}}
\newtheorem{claim}{{\bf Claim}}

\newtheorem{corollary}{{\bf Corollary}}

\newtheorem{remark}{{\bf Remark}}

\newtheorem{defn}{\bf Definition}

\newcommand{\barr}{\begin{array}}
	\newcommand{\earr}{\end{array}}

\newcommand{\benum}{\begin{enumerate}}
	\newcommand{\eenum}{\end{enumerate}}

\newcommand{\bit}{\begin{itemize}}
	\newcommand{\eit}{\end{itemize}}

\newcommand{\bdes}{\begin{description}}
	\newcommand{\edes}{\end{description}}

\newcommand{\bfig}{\begin{figure}}
	\newcommand{\efig}{\end{figure}}

\newcommand{\bemq}{\begin{quote} \begin{em}}
		\newcommand{\eemq}{\end{em} \end{quote}}

\newcommand{\cbrac}[1]{\left\{{#1}\right\}}

\newcommand{\ol}[1]{\overline{#1}}

\newcommand{\indic}[1]{I_{\cbrac{#1}}}

\newcommand{\given}{\arrowvert}

\newcommand{\ie}{{i.e.}}

\newcommand{\expect}[1]{\mathbb{E}\left[{#1}\right]}

\newcommand{\bt}{\begin{theorem}}
	\newcommand{\bl}{\begin{lemma}}
		\newcommand{\bc}{\begin{claim}}
			\newcommand{\bp}{\begin{Proposition}}
				\newcommand{\bcoro}{\begin{corollary}}
					\newcommand{\bres}{\begin{Result}}
						\newcommand{\brem}{\begin{Remark}}
							\newcommand{\et}{\end{theorem}}
						\newcommand{\el}{\end{lemma}}
					\newcommand{\ec}{\end{claim}}
				\newcommand{\ep}{\end{Proposition}}
			\newcommand{\ecoro}{\end{corollary}}
		\newcommand{\eres}{\end{Result}}
	\newcommand{\erem}{\end{Remark}}

\newcommand{\beq}{\begin{equation}}
	\newcommand{\eeq}{\end{equation}}

\newcommand{\norm}[1]{\|{#1}\|}

\newcommand{\mb}[1]{\mathbb{#1}}
\newcommand{\mf}[1]{\mathbf{#1}}
\newcommand{\mc}[1]{\mathcal{#1}}

\begin{document} 
\title{A Throughput Optimal Scheduling Policy for a Quantum Switch}

\author{{\IEEEauthorblockN{1.~Thirupathaiah Vasantam}\\
	\IEEEauthorblockA{\textit{dept. of Computer Science} \\
		\textit{Durham University}\\
		Durham, UK \\
		thirupathaiah.vasantam@durham.ac.uk}}
	\and\\
{\IEEEauthorblockN{2.~Don Towsley}\\
		\IEEEauthorblockA{\textit{College of Information and Computer Sciences} \\
			\textit{University of Massachusetts}\\
			Amherst, USA \\
			towsley@cs.umass.edu}}}



 \maketitle

\begin{abstract}
		We study a quantum switch that creates shared end-to-end entangled quantum states to multiple sets of users that are connected to it. Each user is connected to the switch via an optical link across which bipartite Bell-state entangled states are generated in each time-slot with certain probabilities, and the switch merges entanglements of links to create end-to-end entanglements for users. One qubit of an entanglement of a link is stored at the switch and the other qubit of the entanglement is stored at the user corresponding to the link.  Assuming that qubits of entanglements of links decipher after one time-slot, we characterize the capacity region, which is defined as the set of arrival rates of requests for end-to-end entanglements for which there exists a scheduling policy that stabilizes the switch. We propose a Max-Weight scheduling policy and show that it stabilizes the switch for all arrival rates that lie in the capacity region. We also provide numerical results to support our analysis.
	\end{abstract}

\section{INTRODUCTION}
\label{sec:intro}  

Quantum entanglement is a key component of quantum information systems that enables applications like quantum key distribution (QKD)\cite{Bennett2014,Ekert}, quantum sensing\cite{Eldredge} (e.g., multipartite entanglement for quantum metrology \cite{Giovannetti_2011,Xia:2019jil}), and distributed quantum computing\cite{Broadbent}. These applications motivate the need for a distributed infrastructure (quantum network) that will supply high quality (fidelity) bipartite and multipartite entanglement to end groups of users \cite{Pirandola_2019,Pant,Dahlberg_2019,Van_meter,Bhaskar_2020}; a quantum network consists of a collection of quantum switches connected to each other through optical links. Although several network architectures have been proposed to provide high entanglement rates at high fidelity \cite{lee2020quantum,Ruoyu,Armstrong_2012,Herbauts:13,Hall}, there is still a long road ahead in designing efficient resource allocation algorithms and their performance analysis that can guide us to implement quantum networks at full-scale in future.

In this paper we focus on design and performance analysis of efficient resource allocation algorithms for a single quantum switch that serves incoming requests for end-to-end entanglements to $M$ different groups of users, under the assumption that the switch is connected to $K$ users.
Each user is connected to the switch via a link across which Bell-pairs are generated between user and the switch, and each of the two nodes of a link stores one qubit of an entanglement of the link in quantum memories. When enough Bell-pairs are available at the links corresponding to a group of users, the switch performs a multi-qubit measurement to provide an end-to-end entanglement to the user group. If the switch has to connect two links, it uses Bell-state measurements (BSMs) and when it must connect three or more links, it uses Greenberger-Horne-Zeilinger (GHZ) basis measurements \cite{nielsen00}.

We consider a time-slotted system where requests arrive according to a stochastic process.  Within each type, requests are stored in an infinite capacity queue and processed according to First-Come-First-Served (FCFS) service discipline. In each time-slot, every link creates at most one entanglement, which decoheres after one time-slot\cite{Boxi_li}. Hence, at most one Bell-pair is available at each link in each time-slot to serve requests. Although the expectation is that eventually quantum networks will include switches with many long coherence time quantum memories, this will not be the case in the near term.  For example, first generation quantum networks are likely to use controllable optical delay line buffers \cite{Burmeister:08} to store single qubit at a time. 

The main objective of the switch is to allocate available Bell-pairs in each time-slot cleverly to various requests so that they are processed as quickly as possible. We ask the following research question, what is the capacity region \ie, the set of arrival rates for which there exists
a scheduling policy under which the Markov chain associated with queues of requests have a stationary probability distribution with finite average waiting times of requests? Can we design a scheduling policy that stabilizes the switch for all the arrival rates that belong to the capacity region? In this paper, we address these questions by characterizing this capacity region  and then proposing and analyzing a Max-Weight scheduling policy that stabilizes the switch for all the arrival rates that lie in the capacity region.

\textit{Related work:}
A simple quantum network that connects two users by a series of repeaters was studied in the literature\cite{Shchukin}. The focus was to compute the expected waiting time required to create an end-to-end entanglement across a path with $n$ links, under the assumption that each link creates a link-level entanglement with certain probability and measurement operations are successful probabilistically. The analysis uses Markov chain theory to compute the waiting times of requests, but closed-form expressions were only derived for networks with at most four segments.

The analysis of a single quantum switch connected to several users was investigated in previous works\cite{Gayane1,Gayane2}. First, the rate at which a switch creates bipartite and tripartite entanglements was analyzed, under the assumption that it has capabilities to store one qubit and two qubits per each link\cite{Gayane1}. Later, the analysis was extended to study the switch that generates end-to-end $n$-partite entanglements\cite{Gayane2}. Using Lyapunov stability theory of Markov chains, it was proved that the switch is stable if and only if the number of attached links, $K$, is greater than or equal to $n$. Linear quantum networks with multiplexing capabilities have been studied in the literature\cite{Guha_repeater,dhara2021subexponential,dhara2021multiplexed} to improve end-to-end entanglement generation rates. Quantum networks could be implemented on several physical platforms, the implementation of quantum networks with multiplexing capabilities on dual-species trapped-ion systems was investigated in previous works \cite{dhara2021multiplexed}. A major drawback of previous works \cite{Gayane1,Gayane2,Guha_repeater,dhara2021multiplexed,dhara2021subexponential} is that, entangled states for users were created whenever there are enough link-level entanglements available across links, but they did not consider queues that store  requests that are  waiting for their service. In our modeling and analysis of the switch, similar to previous works\cite{Guha_repeater,dhara2021multiplexed,dhara2021subexponential}, we associate probabilities with various stochastic operations that affect how a quantum switch operates.

A Max-Weight scheduling policy was first introduced for resource allocation in communication networks\cite{Leandros_maxweight} and later, this policy was adopted for the analysis of a single switch in classical networking\cite{McKeown} where they showed that the switch is stable for all feasible arrival rates under this policy. Although the Max-Weight policy has high implementation cost\cite{McKeown}, it lead to a significant progress on design and analysis of low complexity efficient scheduling algorithms in classical networking \cite{srikant_book}. A major challenge in analyzing quantum networks is that they are more dynamic than classical networks due to the fact that several required operations to create end-to-end entanglements are probabilistic operations. Hence, both the design and analysis of scheduling policies must be modified to consider various characteristic properties of quantum networks. For example, if qubits of link-level entanglements decipher after multiple time-slots then the analysis of scheduling policies involves study of two-sided queues, in that one set of queues are used to store requests and the other set of queues are used to store qubits of link-level entanglements; analyzing two-sided queues is very difficult and they are often not needed to study classical networking problems. In this paper, we assume that qubits of entanglements decohere after one time-slot, to simplify the analysis.
In a different context, a Max-Weight scheduling policy that is similar to ours was studied for networks with certain dynamic properties\cite{Leandros_Varying,Kavita_maxweight}. Our analysis is similar in spirit to the Lyapunov stability theory of Markov chains used in these works\cite{Leandros_Varying}.

\textit{Our Contributions:} We make the following contributions:\bit
\item We derive necessary conditions on the request arrival rates for existence of a scheduling policy that stabilizes the switch.
\item We propose a Max-Weight scheduling policy as a function of probability of successful creation of link-level entanglements and measurement operations, and dynamic queue sizes of requests. We prove that this policy stabilizes the switch for all feasible arrival rates using Lyapunov stability theory of Markov chains.
\item Finally, we provide numerical results that corroborate our analysis.
\eit

The rest of the paper is organized as follows.
In Section~\ref{sec:model}, we give details of the system model and then we give notation and some preliminary results in Section~\ref{sec:preliminary} where we also define our Max-Weight scheduling policy. In Section~\ref{sec:results}, we give necessary conditions on the request arrival rates for the stability of the switch and provide main results. We then discuss some numerical results in Section~\ref{sec:numerics}. Finally, we give proofs of theorems in Section~\ref{sec:proofs} and conclude in Section~\ref{sec:conclusions}.                        
\section{System Model}
	\label{sec:model}
Consider a quantum switch that is connected to a set of $K$ users, denoted by $\mb{U}\triangleq\{u_1,\cdots,u_K\}$, in a star topology with user $u_i$ connected to the switch via link $l_i\in \mb{L}\triangleq\{l_1,\cdots,l_k\}$. Time is divided into fixed-length time-slots. One Bell-pair is generated across the link $l_i$ in each time-slot with probability $p_i$; no entanglement is generated with probability $1-p_i$. The Bell-pairs generated across links are referred to as link-level entanglements and the goal of the switch is to merge link-level entanglements to form end-to-end entanglements for various sets of users.

If a link-level entanglement is created between a  user and the switch, then the switch stores one qubit of the entanglement in a quantum memory and the other qubit is stored at the user. The qubits of an entanglement are assumed to decohere after one time-slot. That is, if a link-level entanglement is created in time-slot $n$ then that entanglement must be used to form an end-to-end entanglement in time-slot $n$, otherwise the link-level entanglement is considered to be wasted. As a result, the switch uses only one quantum memory to store qubits of entanglements generated across each link.

$M$ types of requests for end-to-end entanglement arrive at the switch. A type $i$ request is served when the switch creates an $n_i$-qubit end-to-end entanglement shared among a set of users denoted by $\mc{R}_i\triangleq\{u_{i1},\cdots,u_{in_i}\}$. Let $\mc{L}_i$ denote the set of links whose Bell-pairs are required to serve a type~$i$ request, and let $\mc{X}_j$ denote the set of types of requests that require a link-level entanglement of link $j$. The process of creating an end-to-end entanglement shared among users $\mc{R}_i$ involves two steps; successful generation of Bell-pairs across links $\mc{L}_i$ followed by a successful entanglement swapping operation performed on the qubits stored at the switch, which is assumed to occur with probability $q_i$.

Let $A_i(n)$ denote the number of type~$i$ requests that arrive in time-slot $n$ and the process $\{A_i(n)\}$ is a stationary process with $\mb{E}[A_i(n)]=\lambda_i$ where $\lambda_i$ is the rate at which type~$i$ requests arrive in each time-slot. The switch stores requests of each type in an infinite capacity queue and processes them in First-Come-First-Served (FCFS) basis. The main aim of the switch is to make scheduling decisions on how to allocate available link-level entanglements to different types of requests so as to serve requests with finite waiting times.

\begin{figure}
		\centering
		\includegraphics[width=0.65\linewidth]{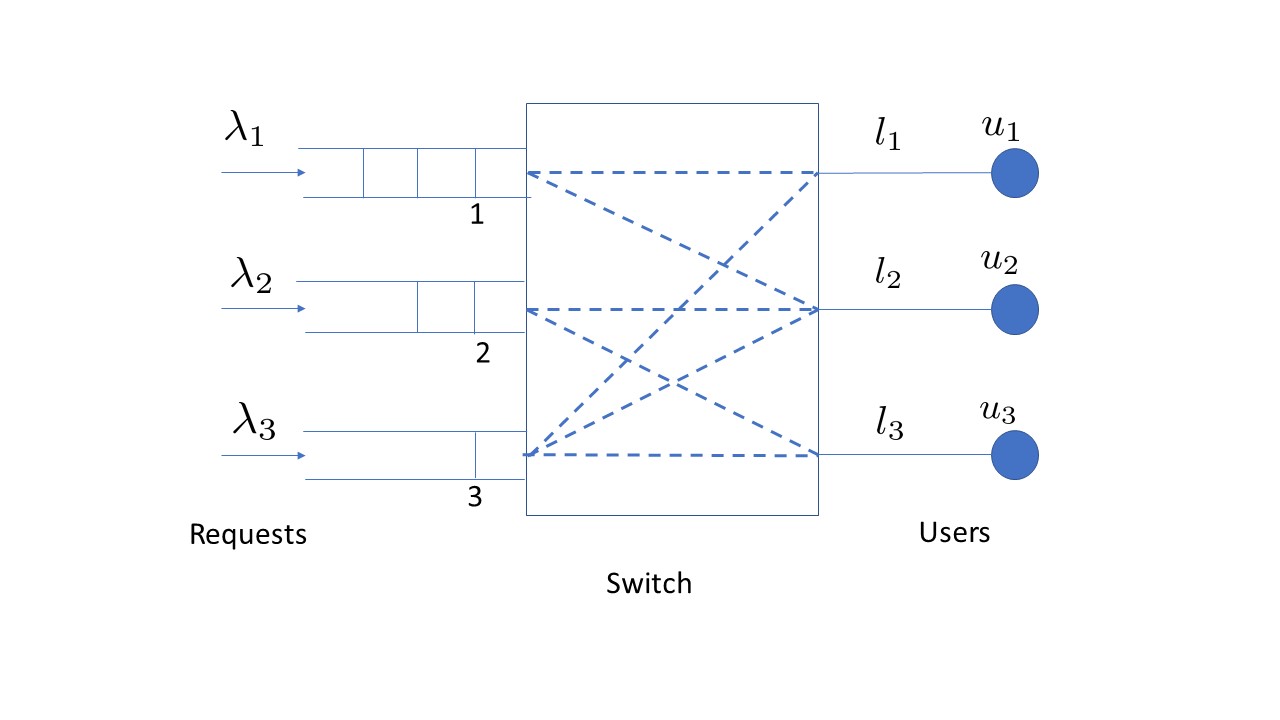}
		\caption{Switch creating end-to-end entanglements}
		\label{fig:switch1}
	\end{figure}

	 In Figure~\ref{fig:switch1}, we show a quantum switch that connects to three users where user $i$ is connected to the switch via link $l_i$. There are three types of requests arriving in the system, each type of request seeks a creation of shared entanglement for a set of users. For type $i$ requests, $\lambda_i$ denotes the average number of requests arriving in each time-slot. 
	 From Figure~\ref{fig:switch1}, the set of users and links associated with different types of requests are: $\mc{R}_1=\{u_1,u_2\}$, $\mc{R}_2=\{u_2,u_3\}$, $\mc{R}_3=\{u_1,u_2,u_3\}$, $\mc{L}_1=\{l_1,l_2\}$, $\mc{L}_2=\{l_2,l_3\}$, and $\mc{L}_3=\{l_1,l_2,l_3\}$. There is a competition among different types of requests to use available link-level entanglements in each time-slot. For example, entanglements of link $l_1$ are needed to serve both type~$1$ and type~3 requests.
	
	In this paper, we address the following question. For a switch with $K$ links and given system parameters $\mf{p}=[p_1,\cdots,p_K]$ and $\mf{q}=[q_1,\cdots,q_M]$, what is the capacity region of request rates that is defined as the set of request rates $\bm{\lambda}=[\lambda_1, \cdots,\lambda_M]$ for which there exists a scheduling policy that stabilizes the switch?	In the next section, we define a Max-Weight scheduling policy, which is shown to stabilize the switch for all feasible arrival rates.

	\section{Notation and Preliminary Results}
	\label{sec:preliminary}
	We write vectors as bold-faced letters in the rest of the paper. We denote the number of type~$i$ requests that are waiting for service at the beginning of time-slot $n$ by $Q_i(n)$. The number of link-level entanglements generated across link $j$ in time-slot $n$ is written as $T_j(n)$, where, $T_j(n)=1$ with probability $p_j$ and $T_j(n)=0$ with probability $1-p_j$. The states of the variable $\mf{T}(n)=(T_j(n))$ belong to the set $\mc{A}$, defined as
	\beq
	\nonumber
	\mc{A}\triangleq
\{\mf{a}=(a_1,\cdots,a_K):
a_i\in\{0,1\}, 1\leq i\leq K\}.
\eeq
 
 In time-slot $n$, since qubits of link-level entanglements decohere after one time-slot, the number of Bell-pairs available at link $j$ is $T_j(n)$. When $T_j(n)=1$, only one request can use it. The switch needs to decide how to process various requests using available link-level entanglements of links, in such a way that the average waiting times of requests are finite. Next we define a notion called matching, that is used in the process of allocating available link-level entanglements to requests. 
	\begin{defn}{Matching:}
		\label{defn:matching}
		We call $\bm{\pi}=[\pi_{1},\cdots,\pi_M]$ a matching if $\pi_{i}\in\{0,1\}$ and for each link $l_j$ ($1\leq j\leq K$), we have
		\beq
		\label{eq:matching}
		\sum_{i\in \mc{X}_j}\pi_{i}\leq 1.
		\eeq
		Furthermore, the vector $\bm{\pi}$ satisfies a condition that if $\pi_r=0$ ($1\leq r\leq M$) in $\bm{\pi}$, then the vector $\bm{\pi}^*$ obtained from $\bm{\pi}$ by replacing the $r^{\text{th}}$ element $\pi_r=0$ with $\pi_r=1$ violates the condition \eqref{eq:matching}. 
	\end{defn}
	
	Condition \eqref{eq:matching} guarantees that each link-level entanglement is assigned to at most one request. If the scheduler selects a matching $\bm{\pi}$ to decide which requests to be served in a time-slot, if $\pi_{i}=1$, then the switch attempts to serve a type~$i$ request by performing a swapping operation on qubits of related links $\mc{L}_i$. 
	Let $\mc{M}$ be the set of all matchings defined as
	\beq
	\nonumber
	\mc{M}\triangleq\{\bm{\pi}: \pi_i\in\{0,1\},\,\sum_{i\in \mc{X}_j}\pi_{i}\leq 1, \forall l_j\}.
	\eeq.

	%
	In time-slot $n$, the switch select a matching from the set $\mc{M}$ based on $\mf{Q}(n)=(Q_{i}(n))$ and $\mf{T}(n)=(T_i(n))$. Intuitively, the switch should allocate available link-level entanglements to types of requests that have large queues, which guides us to define a Max-Weight scheduling policy for given quantum switch. Suppose that $\mf{W}(n)\in\mc{M}$ is the matching to be used in time-slot $n$, then we denote $r_i(\mf{T}(n),\mf{W}(n))$ to be the probability that a type~$i$ request is successfully served given that it is selected for service. To serve a type~$i$ request, first, the switch should make a decision to perform a relevant swapping operation, which happens if all the links in $\mc{L}_i$ have Bell-pairs and the selected matching $\mf{W}(n)$ satisfies $W_i(n)=1$. Second, the subsequent swapping operation must succeed, which happens with probability $q_i$. As a result,
	\beq
	\label{eq:success_probability}
	r_i(\mf{T}(n),\mf{W}(n))=q_i\indic{W_i(n)=1}\indic{T_j(n)>0,\forall l_j\in \mc{L}_i},
	\eeq
	where, $\indic{B}$ is the indicator function of the event $B$.
	
	Next, we define the Max-Weight scheduling policy of interest below.
	\begin{defn}{Max-Weight Scheduling:}
		In time-slot $n$, the switch selects the matching $\mf{W}(n)$ computed as follows:
		\beq
		\label{eq:max_weight}
		\mf{W}(n)=\arg\max_{\bm{\pi}\in\mc{M}}\sum_{i=1}^Mr_i(\mf{T}(n),\bm{\pi})Q_{i}(n).
		\eeq
		\end{defn}	
	From \eqref{eq:max_weight}, it is clear that $\mf{W}(n)$ is chosen to maximize the weighted sum of queue sizes of requests with weights corresponding to success probabilities of serving requests over the set $\mc{M}$. This helps us to avoid congested queues.



	Next, we show how the process $\{\mf{Q}(n)\}$ evolves with time. Suppose that $Z_i(n)\in\{0,1\}$ denotes whether an entanglement swapping operation performed on qubits of links $\mc{L}_i$ in time-slot $n$ succeeds or not. Variable $\mf{Z}(n)$ satisfies $Z_i(n)=1$ if the entanglement swapping operation succeeds and $Z_i(n)=0$, otherwise.
 Now define $D_i(n)\in\{0,1\}$ to be the number of type~$i $requests served in time-slot $n$. Then we have
	\beq
	\label{eq:departure}
	D_i(n)=Z_i(n)\indic{W_i(n)>0}\indic{Q_i(n)>0}\indic{T_j(n)>0,\forall l_j\in \mc{L}_i}. 
	\eeq
	Process $\{\mf{Q}(n)\}$	is a Markov chain that evolves according to the following relation,
	\beq
	\label{eq:queue_dynamics}
	\mf{Q}(n+1)=\mf{Q}(n)-\mf{D}(n)+\mf{A}(n),
	\eeq
	where, $\mf{A}(n)=(A_i(n))$ and $\mf{D}(n)=(D_i(n))$.
	Note that the newly arrived requests $\mf{A}(n)$	are not used to compute $\mf{D}(n)$, but rather are used to determine $\mf{D}(n+1)$. 

	Our goal is to find necessary conditions on $\bm{\lambda}$ for existence of a scheduling policy under which the switch is stable, that there exists a stationary probability distribution for queue sizes of requests with finite average queues. We will show that our Max-Weight policy, as defined below, stabilizes the switch for all arrival rates belonging to the capacity region. 
	\begin{defn}{Capacity Region:}
	The set of request rates $\bm{\lambda}$ for which there exists a scheduling policy that stabilizes the switch.
	\end{defn}
	
		A scheduling policy is said to be throughput optimal if it stabilizes the switch for all arrival rates belonging to the capacity region. In the following remark, we will recall results on the analysis of a switch in classical networking, and then discuss how classical and quantum switch differ in the way they operate. 
	\begin{remark}
		In classical networking, a switch forwards packets from input ports to output ports, under the condition that in each time-slot, an input port forwards at most one packet to only one output port and an output port receives at most one packet from only one input port.
		 Suppose that $\lambda_{ij}$ denotes the average number of arriving packets per time-slot at the input port $i$ to be transferred to the output port $j$.  Define $\Lambda'$ as
		\beq
		\nonumber
		\Lambda'=\{\mf{a}=[a_{ij}]:\sum_{j}a_{ij}\leq 1\text{ and }\sum_{l}a_{lm}\leq 1,\, \forall\,i,m\}.
		\eeq
		Let $\mc{M}'$ denote the set of matchings used in classical networking defined as
		\beq
		\nonumber
		\mc{M}'\triangleq\{\bm{\pi}=[\pi_{ij}]:\sum_{j}\pi_{ij}= 1\text{ and }\sum_{l}\pi_{lm}= 1,\, \forall\,i,m\}.
		\eeq
		In \cite{McKeown}, it was shown that if the switch selects the matching $W(n)$ computed according to the following Max-Weight scheduling policy then the switch is stable if $\bm{\lambda}$ lies inside $\Lambda'$,
		\beq
		\nonumber
		\label{eq:max_weight2}
		\mf{W}(n)=\arg\max_{\pi\in\mc{M}'}\sum_{ij}\pi_{ij}Q_{ij}(n).
		\eeq
		Furthermore, if $\bm{\lambda}\notin \Lambda'$, then no scheduling policy can stabilize the switch. We can view the quantum switch as the device with $M$ input ports and $K$ output ports, where each input port is associated with an application that generates requests for end-to-end entanglements and each output port is associated with a link. In every time-slot, the input port $i$ is either matched to output ports $\mc{L}_i$ or not matched to any output port depending on whether $W_i(n)=1$ or not. Furthermore, each output port is matched to  at most one input port since each link has at most one link-level entanglement. If the input port $i$ is matched to output ports, then it means that the switch has decided to serve a type~$i$ request. 	\end{remark}
	
	In the next section, we will derive necessary conditions on $\bm{\lambda}$ for achieving the stability of the switch and show that the proposed Max-Weight scheduling policy achieves the stability of the switch for all arrival rates that lie inside the capacity region.

\section{Main Results}
\label{sec:results}
In this section, we present necessary conditions on arrival rates $\bm{\lambda}$ to achieve stability of the switch and prove that the Max-Weight policy stabilizes the switch for all feasible arrival rates using Lyapunov stability theory of Markov chains.

Next we provide a formal definition of stability of the switch.
\begin{defn}{Stability of the switch:}
We say that the quantum switch is stable if the sequence $\{\mf{Q}(n)\}$ converges in distribution to a random vector $\mf{Q}(\infty)$ with $\expect{\mf{Q}(\infty)}<\infty$ for all initial states $\mf{Q}(0)$.
\end{defn}

 In our proofs we use the condition that the process $\{\mf{Q}(n)\}$ is an irreducible Markov chain. The process $\{\mf{Q}(n)\}$ is an irreducible Markov chain under a scheduling policy if the following two conditions are satisfied. These are:
\begin{description}
\item{$C_1:$} If $\lambda_i>0$, then there exists $\bm{\pi}\in\mc{M}$ such that $r_i(\mf{T}(n),\bm{\pi})>0$ for some $\mf{T}(n)$.  
\item{ $C_2:$} If $\mf{Q}(n)\neq 0$, then there exists a matching $\bm{\pi}$ such that $r_i(\mf{T}(n),\bm{\pi})>0$ for a given state of $\mf{T}(n)$ for some $i$ with $Q_i(n)>0$, in this case the scheduling policy of interest must select a matching $\mf{W}(n)$ such that 
$r_j(\mf{T}(n),\mf{W}(n))>0$ for some $j$ with $Q_j(n)>0$.
\end{description}
If a scheduling policy satisfies conditions $C_1$ and $C_2$, then the process $\{\mf{Q}(n)\}$ is guaranteed to reach the empty state starting from any initial state.
From \eqref{eq:success_probability} and \eqref{eq:max_weight}, it is evident that the two conditions, $C_1$ and $C_2$, are satisfied under our Max-Weight scheduling policy. Hence, the process $\{\mf{Q}(n)\}$ is an irreducible Markov chain.

If the switch is stable under a scheduling policy then the request arrival rate coincides with the request departure rate, 
\beq
\bm{\lambda}=\lim_{n\to\infty}\frac{\sum_{j=1}^n\mf{D}(j)}{n}, \quad\quad a.s.
\eeq
 
 Next, we derive necessary conditions on $\bm{\lambda}$ that guarantee that the switch can be stabilized under a scheduling policy. If the switch is stable under a scheduling policy, then we denote $\mf{X}(\infty)$ to be the random vector with stationary probability distribution of $\mf{X}(n)$. Let $c_{a,\bm{\pi}}$ be defined as
 \beq
 \nonumber
 c_{a,\bm{\pi}}=\mb{P}(\min(\mf{W}(\infty),\mf{Q}(\infty))=\bm{\pi},\,\mf{Q}(\infty)\neq\mf{0}\given \mf{T}(\infty)=\mf{a}),
 \eeq
 where $\min(\mf{W}(\infty),\mf{Q}(\infty))=(\min(W_i(\infty),Q_i(\infty)))$ indicates the number of entanglement swapping operations performed for each type of requests in time-slot $n$, and  $c_{a,\bm{\pi}}$ denotes the stationary probability that $\min(\mf{W}(\infty),\mf{Q}(\infty))=\bm{\pi}$ and $\mf{Q}(\infty)\neq\mf{0}$ given that $\mf{T}(\infty))=\mf{a}$. Note that the process $\{\mf{T}(n)\}$ is an i.i.d. process with the property that $\mb{P}(\mf{T}(n)=1)=p_i$ and $\mb{P}(\mf{T}(n)=0)=1-p_i$.
 \begin{theorem}
 \label{thm:arr_stability_condn}
 If the switch is stable under a scheduling policy and $\{\mf{Q}(n)\}$ is an irreducible Markov chain, then $\bm{\lambda}$ satisfies
 \beq
\lambda
=\sum_{\{\mf{a}\in\mc{A},\mf{a}\neq \mf{0}\}}\mb{P}(\mf{T}(n)=\mf{a})
\sum_{\bm{\pi}\in\mc{M}}c_{\mf{a},\bm{\pi}}\mf{r}(\mf{a},\bm{\pi}),\label{eq:lambda_nec_condn}
\eeq
where, $\mf{r}(\mf{a},\bm{\pi})=(r_i(\mf{a},\bm{\pi}))$,  $c_{\mf{a},\bm{\pi}}>0$, and $\sum_{\bm{\pi}}c_{\mf{a},\bm{\pi}}<1$ for all $\mf{a}\in\mc{A}$ with $\mf{a}\neq\mf{0}$.
\end{theorem}
\begin{proof}
The proof is given in Section~\ref{proof:arr_stability_condn}.
\end{proof}

From \eqref{eq:lambda_nec_condn}, we can write
\begin{align}
\label{eq:new_lambda_condn}
\lambda
&=\sum_{\{\mf{a}\in\mc{A},\mf{a}\neq \mf{0}\}}\mb{P}(\mf{T}(n)=\mf{a})
\sum_{\bm{\sigma}\in\mc{M}}\sum_{\bm{\pi}\in\mc{M}}\mb{P}(\mf{W}(\infty)=\bm{\sigma},\min(\mf{W}(\infty),\mf{Q}(\infty))=\bm{\pi},\,\mf{Q}(\infty)\neq\mf{0}\given \mf{T}(\infty)=\mf{a})\mf{r}(\mf{a},\bm{\pi})\nonumber\\
&\leq\sum_{\{\mf{a}\in\mc{A},\mf{a}\neq \mf{0}\}}\mb{P}(\mf{T}(n)=\mf{a})
\sum_{\bm{\sigma}\in\mc{M}}\sum_{\bm{\pi}\in\mc{M}}\mb{P}(\mf{W}(\infty)=\bm{\sigma},\min(\mf{W}(\infty),\mf{Q}(\infty))=\bm{\pi},\,\mf{Q}(\infty)\neq\mf{0}\given \mf{T}(\infty)=\mf{a})\mf{r}(\mf{a},\bm{\sigma})\nonumber\\
&=\sum_{\{\mf{a}\in\mc{A},\mf{a}\neq \mf{0}\}}\mb{P}(\mf{T}(n)=\mf{a})
\sum_{\bm{\sigma}\in\mc{M}}\mb{P}(\mf{W}(\infty)=\bm{\sigma},\,\mf{Q}(\infty)\neq\mf{0}\given \mf{T}(\infty)=\mf{a})\mf{r}(\mf{a},\bm{\sigma})\nonumber\\
&=\sum_{\{\mf{a}\in\mc{A},\mf{a}\neq \mf{0}\}}\mb{P}(\mf{T}(n)=\mf{a})
\sum_{\bm{\sigma}\in\mc{M}}b_{\mf{a},\bm{\sigma}}\mf{r}(\mf{a},\bm{\sigma}),
\end{align}
where $b_{\mf{a},\bm{\sigma}}=\mb{P}(\mf{W}(\infty)=\bm{\sigma},\,\mf{Q}(\infty)\neq\mf{0}\given \mf{T}(\infty)=\mf{a})$.

Using Theorem~\ref{thm:arr_stability_condn}, we characterize the capacity region as follows.
\begin{defn}{Capacity region:}
The capacity region is defined as
\begin{multline}
\Lambda\triangleq\Bigg\{\bm{\lambda}: \exists \{  b_{\mf{a},\bm{\pi}},\mf{a}\in\mc{A},\bm{\pi}\in{\mc{M}}\}\text{  such that }\\
\bm{\lambda}
\leq\sum_{\{\mf{a}\in\mb{A},\mf{a}\neq \mf{0}\}}\mb{P}(\mf{T}(n)=\mf{a})
\sum_{\bm{\pi}}b_{\mf{a},\bm{\pi}}\mf{r}(\mf{a},\bm{\pi}),\, b_{\mf{a},\bm{\pi}}>0,\,\sum_{\bm{\pi}}b_{\mf{a},\bm{\pi}}<1,\forall\mf{a}\Bigg\}.
\end{multline}
\end{defn}
If $\lambda\notin\bm{\Lambda}$, then the switch cannot be stabilized under any scheduling policy as it would contradict the results of Theorem~\ref{thm:arr_stability_condn}. 

Next, we prove that the Max-Weight scheduling policy stabilizes the switch for all arrival rates in the capacity region. For this, we apply a Lyapunov stability theorem of Markov chains \cite[Theorem~3.1]{Leandros_maxweight}, using the following Lyapunov function
\[
V(\mf{Q}(n))=\sum_{i=1}^MQ_i(n)^2.
\]
It suffices to show that
\beq 
\label{eq:drift_condition}
\expect{V(\mf{Q}(n+1))-V(\mf{Q}(n))\given \mf{Q}(n)}\leq -\epsilon \norm{\mf{Q}(n)},
\eeq
for sufficiently large $\norm{\mf{Q}(n)}$, where $\norm{\mf{Q}(n)}=\sqrt{\sum_{i=1}^MQ_i(n)^2}$, and $\epsilon>0$. 
Finally, we state the main result on the stability of the switch under our Max-Weight scheduling policy in the following theorem.
\begin{theorem}
\label{thm:stability_result}
If $\bm{\lambda}\in\Lambda$ and $\expect{A_i^2(n)}<\infty$ for all $1\leq i\leq M$, then the Max-Weight scheduling policy defined in Definition~\ref{defn:matching} stabilizes the switch.
\end{theorem}
\begin{proof}
The proof is given in Section~\ref{proof:stability_result}.
\end{proof}

\section{Numerical Results}
\label{sec:numerics}
In this section, we provide numerical results that support our analysis. We simulate the switch shown in Figure~\ref{fig:switch1} to understand the behavior of the process $\{\ol{\mf{Q}}(n)\}$ for various parameters, where $\ol{\mf{Q}}(n)=\frac{\sum_{i=1}^MQ_i(n)}{M}$.

\begin{figure}
		\centering
		\includegraphics[width=0.55\linewidth]{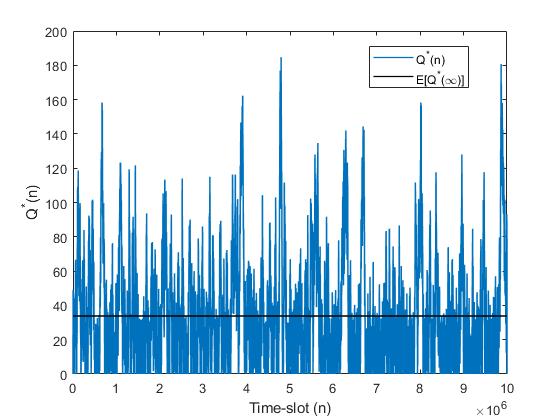}
		\caption{Evolution of queue sizes in a stable switch}
		\label{fig:stable_switch}
	\end{figure}

\begin{figure}
		\centering
		\includegraphics[width=0.55\linewidth]{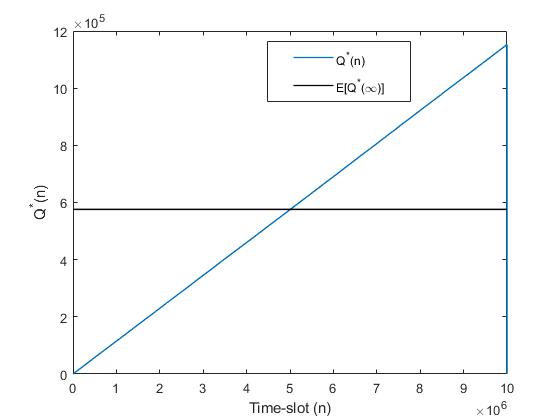}
		\caption{Evolution of queue sizes in an unstable switch}
		\label{fig:unstable_switch}
	\end{figure}

 First, we consider parameters $\lambda=$[0.35 0.2 0.15], $\mf{p}=$[0.7 0.8 0.6], and $\mf{q}=$[0.9 0.8 0.7]. In Figure~\ref{fig:stable_switch}, we plot $\ol{\mf{Q}}(n)$ as a function of $n$. This figure shows that switch is stable for the considered parameters and the stationary average queue size denoted by $\mb{E}[{\ol{\mf{Q}}(\infty)}]$ is finite as shown in the figure, where $\mb{E}[\ol{\mf{Q}}(\infty)]=\frac{\sum_{n=1}^N\ol{\mf{Q}}(n)}{N}$ with $N=10^7$. In Figure~\ref{fig:unstable_switch}, we study the switch assuming higher request rates than the arrival rates considered in Figure~\ref{fig:stable_switch}. For $\mf{\lambda}=$[0.45 0.35 0.25], we observe that $\ol{\mf{Q}}(n)$ increases monotonically with $n$ as shown in Figure~\ref{fig:unstable_switch} implying that the switch is unstable and $\mb{E}[\ol{\mf{Q}}(\infty)]$ is very large. 
 
\begin{figure}
		\centering
		\includegraphics[width=0.55\linewidth]{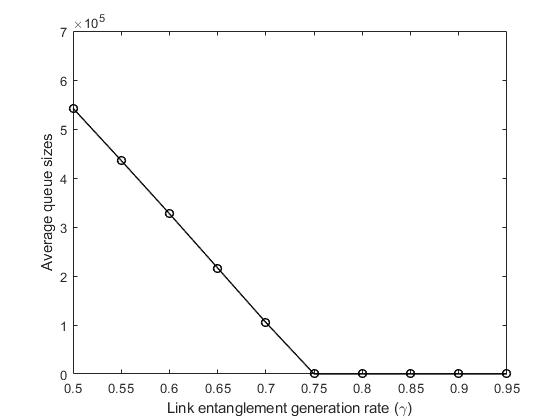}
		\caption{Average queue sizes versus $\gamma$}
		\label{fig:unstable_switch_gamma}
	\end{figure}

\begin{figure}
		\centering
		\includegraphics[width=0.55\linewidth]{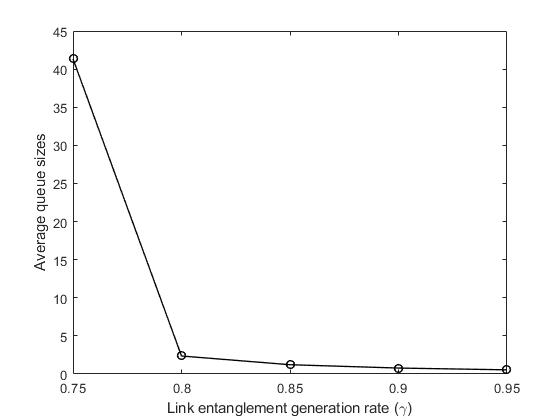}
		\caption{Average queue sizes versus $\gamma$ in a stable switch}
		\label{fig:stable_switch_gamma}
	\end{figure}

 In Figure~\ref{fig:unstable_switch_gamma}, we study the switch with parameters $\lambda=$[0.35 0.2 0.15], $\mf{p}=$[$\gamma$ $\gamma$ $\gamma$], and $\mf{q}=$[0.9 0.8 0.7]. We plot $\mb{E}[\ol{\mf{Q}}(\infty)]$ as a function of $\gamma$ in Figure~\ref{fig:unstable_switch_gamma}. We observe that the switch is unstable when $\gamma<0.75$ due to the fact that there are not enough link-level entanglements available in each time-slot to serve requests stored in queues. The average queue sizes of requests decrease with link-level entanglement generation rate $\gamma$. In Figure~\ref{fig:stable_switch_gamma}, for $\gamma\geq 0.75$, we observe that the average queue sizes of requests are small and decrease with $\gamma$. Our numerical results support the importance of characterizing the capacity region of the switch for given $\mf{p}$ and $\mf{q}$.

	

\section{Proofs}
\label{sec:proofs}
\subsection{Proof of Theorem~\ref{thm:arr_stability_condn}}
	\label{proof:arr_stability_condn}
Since the process $\{\mf{Q}(n)\}$ is an irreducible Markov chain and the process $\{\mf{T}(n)\}$ is an i.i.d. process with finite states, the process $\{\mf{Q}(n),\mf{T}(n)\}$ is also an irreducible Markov chain. Assume that the switch begins with an initial state having the stationary distribution, then in that case the process $\{\mf{Q}(n),\mf{T}(n)\}$ is a stationary process. At steady state, we have
\begin{align}
\label{eq:rate_equality}
\lambda&=\mb{E}(\mf{D}(n))  \nonumber \\
&=\mb{E}\left(\mb{E}[\mf{D}(n)\given \mf{Q}(n),\mf{T}(n)]\right).
\end{align}
 We can write \eqref{eq:rate_equality} as
\beq
\lambda
=\sum_{\{\mf{a}\in\mc{A},\mf{a}\neq \mf{0}\}}\mb{P}(\mf{T}(n)=\mf{a})\sum_{\{\mf{b}\in\mb{Z}_+^d,\mf{b}\neq \mf{0}\}}\mb{P}(\mf{Q}(n)=\mf{b})\mb{E}[\mf{D}(n)\given \mf{Q}(n)=\mf{b},\mf{T}(n)=\mf{a}],
\eeq
where, $\mb{Z}_+$ denotes the set of non-negative integers. In the above equation, we have also used the fact that $\mf{D}(n)=\mf{0}$ if $\mf{Q}(n)=\mf{0}$ or $\mf{T}(n)=\mf{0}$. From \eqref{eq:departure}, we obtain
\beq
\label{eq:new_arr}
\lambda
=\sum_{\{\mf{a}\in\mc{A},\mf{a}\neq \mf{0}\}}\mb{P}(\mf{T}(n)=\mf{a})\sum_{\{\mf{b}\in\mb{Z}_+^d,\mf{b}\neq \mf{0}\}}\mb{P}(\mf{Q}(n)=\mf{b})
\left\{\sum_{\bm{\pi}\in\mc{M}}\mb{P}(\mf{W}(n)=\bm{\pi}\given \mf{T}(n)=\mf{a},\mf{Q}(n)=\mf{b})\mf{r}(\mf{a},\min(\mf{b},\bm{\pi}))\right\},
\eeq
where $\min(\mf{b},\bm{\pi}))=(\min(b_i,\pi_i))$. Note that we can write $\min(b_i,\pi_i)=\indic{b_i>0}\indic{\pi_i>0}$.
For $\bm{\sigma}=(\sigma_i)$, we can write \eqref{eq:new_arr} as
\beq
\lambda
=\sum_{\{\mf{a}\in\mc{A},\mf{a}\neq \mf{0}\}}\mb{P}(\mf{T}(n)=\mf{a})\sum_{\bm{\sigma}\in\mc{M}}\mb{P}(\min(\mf{W}(n),\mf{Q}(n))=\bm{\sigma}, \mf{Q}(n)\neq \mf{0}\given \mf{T}(n)=\mf{a})\mf{r}(\mf{a},\bm{\sigma}).\nonumber
\eeq

By defining $c_{\mf{a},\bm{\sigma}}$ as
$
c_{\mf{a},\bm{\sigma}}=\mb{P}(\min(\mf{W}(n),\mf{Q}(n))=\bm{\sigma}, \mf{Q}(n)\neq \mf{0}\given \mf{T}(n)=\mf{a}),
$ we obtain
\beq
\lambda
=\sum_{\{\mf{a}\in\mc{A},\mf{a}\neq \mf{0}\}}\mb{P}(\mf{T}(n)=\mf{a})
\sum_{\bm{\sigma}\in\mc{M}}c_{\mf{a},\bm{\sigma}}\mf{r}(\mf{a},\bm{\sigma}).\nonumber
\eeq
Since $\{\mf{Q}(n)\}$ is an irreducible Markov chain, at steady state we have $\mb{P}(\mf{Q}(n)=\mf{0})>0$, it then follows that
$
\sum_{\bm{\sigma}\in\mc{M}}c_{\mf{a},\bm{\sigma}}<1
$
for all $\mf{a}\in\mc{A}$.


\subsection{Proof of Theorem~\ref{thm:stability_result}}
\label{proof:stability_result}
We derive expression for $\expect{V(\mf{Q}(n+1))-V(\mf{Q}(n))\given \mf{Q}(n)}$ and obtain some useful bounds to complete the proof.
 Denote $\mf{a}\cdot\mf{b}$ to be the dot product of two vectors $\mf{a}=(a_i,1\leq i\leq M)$ and $\mf{b}=(b_i,1\leq i\leq M)$, where $\mf{a}\cdot\mf{b}=\sum_{i=1}^Ma_ib_i$. In the following lemma, we obtain a useful result.
\begin{lemma}
\label{thm:second_part}
Under the Max-Weight policy defined in Definition~\ref{defn:matching}, we prove that
\beq
\expect{V(\mf{Q}(n+1))-V(\mf{Q}(n))\given \mf{Q}(n),\mf{T}(n)}\\
\leq \sum_{i=1}^M\expect{A_i(n)^2}+M
+2\left(\bm{\lambda}\cdot\mf{Q}(n) -\mf{r}(\mf{T}(n),\mf{W}(n))\cdot \mf{Q}(n)\right).
\eeq
\end{lemma}
\begin{proof}
 By writing $V(\mf{Q}(n))=\mf{Q}(n)\cdot \mf{Q}(n)$,
we obtain
\beq
V(\mf{Q}(n+1))-V(\mf{Q}(n))=\mf{Q}(n+1) \cdot \mf{Q}(n+1)-\mf{Q}(n) \cdot \mf{Q}(n).\nonumber
\eeq
We can write the above equation as
\beq
V(\mf{Q}(n+1))-V(\mf{Q}(n))
=(\mf{Q}(n+1)-\mf{Q}(n))\cdot (\mf{Q}(n+1)-\mf{Q}(n))
+2((\mf{Q}(n+1)-\mf{Q}(n))\cdot\mf{Q}(n)).\nonumber
\eeq
Using \eqref{eq:queue_dynamics}, we get
\beq
V(\mf{Q}(n+1))-V(\mf{Q}(n))
\leq \mf{A}(n)\cdot\mf{A}(n)+\mf{D}(n)\cdot\mf{D}(n)\\
+2((\mf{Q}(n+1)-\mf{Q}(n))\cdot\mf{Q}(n)).\nonumber
\eeq
Therefore, we obtain
\beq
\label{eq:drift_bound}
\expect{V(\mf{Q}(n+1))-V(\mf{Q}(n))\given \mf{Q}(n),\mf{T}(n)}
\leq \sum_{i=1}^M\expect{A_i(n)^2}+M
+2\expect{(\mf{Q}(n+1)-\mf{Q}(n))\cdot\mf{Q}(n)\given \mf{Q}(n),\mf{T}(n)}.
\eeq

To complete the proof, it remains to simplify the third term on the right side of \eqref{eq:drift_bound}. From  \eqref{eq:queue_dynamics}, we get
\beq
\expect{(\mf{Q}(n+1)-\mf{Q}(n))\cdot\mf{Q}(n)\given \mf{Q}(n),\mf{T}(n)}
=\expect{\mf{A}(n)\cdot\mf{Q}(n)\given \mf{Q}(n),\mf{T}(n)}
-\expect{\mf{D}(n)\cdot\mf{Q}(n)\given \mf{Q}(n),\mf{T}(n)}.\nonumber
\eeq
The above equation can be simplified as
\beq
\expect{(\mf{Q}(n+1)-\mf{Q}(n))\cdot\mf{Q}(n)\given \mf{Q}(n),\mf{T}(n)}
=\bm{\lambda}\cdot\mf{Q}(n) -\mf{r}(\mf{T}(n),\mf{W}(n))\cdot \mf{Q}(n).\nonumber
\eeq
This completes the proof.
\end{proof}

We now establish the condition \eqref{eq:drift_condition} using Lemma~\ref{thm:second_part}.
    From Lemma~~\ref{thm:second_part}, we have
    \begin{multline}
    \label{eq:drift_merge}
    \expect{V(\mf{Q}(n+1))-V(\mf{Q}(n))\given \mf{Q}(n)}\\
    \leq \sum_{i=1}^M\expect{A_i(n)^2}+M
+2\big[\bm{\lambda}\cdot\mf{Q}(n) -\sum_{\{\mf{a}\in\mc{A},\mf{a}\neq \mf{0}\}}\mb{P}(\mf{T}(n)=\mf{a})\max_{\bm{\pi}\in \mc{M}}\{\mf{r}(\mf{a},\bm{\pi})\cdot \mf{Q}(n)\}\big].
    \end{multline}
    Since $\bm{\lambda}\in\Lambda$ there exists $\{  b_{\mf{a},\bm{\sigma}},\mf{a}\in\mc{A},\bm{\sigma}\in{\mc{M}}\}$ such that  $b_{\mf{a},\bm{\sigma}}>0$,\,$\sum_{\bm{\sigma}\in\mc{M}}b_{\mf{a},\bm{\sigma}}<1$ for all $\mf{a}$, and
  \beq 
\lambda
\leq\sum_{\{\mf{a}\in\mc{A},\mf{a}\neq \mf{0}\}}\mb{P}(\mf{T}(n)=\mf{a})
\sum_{\bm{\sigma}\in\mc{M}}b_{\mf{a},\bm{\sigma}}\mf{r}(\mf{a},\bm{\sigma}).\nonumber
\eeq
As a consequence, we have 
\beq
\label{eq:queue_arrival_inner}
\lambda\cdot \mf{Q}(n)
\leq\sum_{\{\mf{a}\in\mc{A},\mf{a}\neq \mf{0}\}}\mb{P}(\mf{T}(n)=\mf{a})
\sum_{\bm{\sigma}\in\mc{M}}b_{\mf{a},\bm{\sigma}}\sum_{i=1}^M r_i(\mf{a},\bm{\sigma}) Q_i(n).
\eeq
Using \eqref{eq:queue_arrival_inner}, we can write
\begin{multline}
\bm{\lambda}\cdot\mf{Q}(n) -\sum_{\{\mf{a}\in\mc{A},\mf{a}\neq \mf{0}\}}\mb{P}(\mf{T}(n)=\mf{a})\max_{\bm{\pi}\in \mc{M}}\left\{\mf{r}(\mf{a},\bm{\pi})\cdot \mf{Q}(n)\right\}\\
\leq\sum_{\{\mf{a}\in\mc{A},\mf{a}\neq \mf{0}\}}\mb{P}(\mf{T}(n)=\mf{a})
\sum_{\bm{\sigma}\in\mc{M}}b_{\mf{a},\bm{\sigma}}( \mf{r}(\mf{a},\bm{\sigma})\cdot\mf{Q}(n)-\max_{\bm{\pi} }\left\{\mf{r}(\mf{a},\bm{\pi})\cdot \mf{Q}(n)\right\})\\
-\sum_{\{\mf{a}\in\mc{A},\mf{a}\neq \mf{0}\}}\mb{P}(\mf{T}(n)=\mf{a})(1-\sum_{\bm{\sigma}}b_{\mf{a},\bm{\sigma}})\max_{\bm{\pi}}\left\{\mf{r}(\mf{a},\bm{\pi})\cdot \mf{Q}(n)\right\}.\nonumber
\end{multline}
Since the first term on the right side of the above equation is negative, we write
\begin{multline}
\bm{\lambda}\cdot\mf{Q}(n) -\sum_{\{\mf{a}\in\mc{A},\mf{a}\neq \mf{0}\}}\mb{P}(\mf{T}(n)=\mf{a})\max_{\bm{\pi}\in \mc{M}}\left\{\mf{r}(\mf{a},\bm{\pi})\cdot \mf{Q}(n)\right\}\\
\leq
-\sum_{\{\mf{a}\in\mc{A},\mf{a}\neq\mf{0}\}}\mb{P}(\mf{T}(n)=\mf{a})(1-\sum_{\bm{\sigma}}b_{\mf{a},\bm{\sigma}})\max_{\bm{\pi}\in \mc{M}}\left\{\mf{r}(\mf{a},\bm{\pi})\cdot \mf{Q}(n)\right\}.\nonumber
\end{multline}
Denote $\delta=\min_{\{\mf{a}\in\mc{A},\mf{a}\neq \mf{0}\}}(1-\sum_{\bm{\sigma}\in\mc{M}}b_{\mf{a},\bm{\sigma}})$, then we have $\delta>0$. We then obtain the following relation
\beq
\bm{\lambda}\cdot\mf{Q}(n) -\sum_{\{\mf{a}\in\mc{A},\mf{a}\neq \mf{0}\}}\mb{P}(\mf{T}(n)=\mf{a})\max_{\bm{\pi}\in \mc{M}}\left\{\mf{r}(\mf{a},\bm{\pi})\cdot \mf{Q}(n)\right\}\\
\leq
-\delta\sum_{\{\mf{a}\in\mc{A},\mf{a}\neq \mf{0}\}}\mb{P}(\mf{T}(n)=\mf{a})\max_{\bm{\pi}\in \mc{M}}\left\{\mf{r}(\mf{a},\bm{\pi})\cdot \mf{Q}(n)\right\}.\nonumber
\eeq
Denote $i_{max}$ as $i_{max}=\arg\max_i Q_i(n)$. Let $\mc{M}^*$ be the set defined as
\beq
\mc{M}^*\triangleq\{\bm{\pi}\in\mc{M}:\pi_{i_{max}}>0\}.\nonumber
\eeq
Then we have
\beq
\max_{\bm{\pi}\in \mc{M}^*}\left\{\mf{r}(\mf{a},\bm{\pi})\cdot \mf{Q}(n)\right\}\leq \max_{\bm{\pi}\in \mc{M}}\left\{\mf{r}(\mf{a},\bm{\pi})\cdot \mf{Q}(n)\right\}.\nonumber
\eeq
As a consequence, we write
\beq
\bm{\lambda}\cdot\mf{Q}(n) -\sum_{\{\mf{a}\in\mc{A},\mf{a}\neq \mf{0}\}}\mb{P}(\mf{T}(n)=\mf{a})\max_{\bm{\pi}\in \mc{M}}\left\{\mf{r}(\mf{a},\bm{\pi})\cdot \mf{Q}(n)\right\}\\
\leq
-\delta\sum_{\{\mf{a}\in\mc{A},\mf{a}\neq \mf{0}\}}\mb{P}(\mf{T}(n)=\mf{a})\max_{\bm{\pi}\in \mc{M}^*}\left\{\mf{r}(\mf{a},\bm{\pi})\cdot \mf{Q}(n)\right\}.\nonumber
\eeq
Since $r_{i_{max}}(\mf{a},\bm{\pi})Q_{i_{max}}(n)\leq \mf{r}(\mf{a},\bm{\pi})\cdot \mf{Q}(n)$, we obtain
\begin{multline}
\bm{\lambda}\cdot\mf{Q}(n) -\sum_{\{\mf{a}\in\mc{A},\mf{a}\neq \mf{0}\}}\mb{P}(\mf{T}(n)=\mf{a})\max_{\bm{\pi}\in \mc{M}}\left\{\mf{r}(\mf{a},\bm{\pi})\cdot \mf{Q}(n)\right\}\\
\leq
-\delta\sum_{\{\mf{a}\in\mc{A},\mf{a}\neq \mf{0}\}}\mb{P}(\mf{T}(n)=\mf{a})\max_{\bm{\pi}\in \mc{M}^*}\left\{r_{i_{max}}(\mf{a},\bm{\pi})Q_{i_{max}}(n)\right\}.\nonumber
\end{multline}
It can be verified that we have
\beq
\nonumber
max_{\bm{\pi}\in \mc{M}^*}\left\{r_{i_{max}}(\mf{a},\bm{\pi})Q_{i_{max}}(n)\right\}
=max_{\bm{\pi}\in \mc{M}}\left\{r_{i_{max}}(\mf{a},\bm{\pi})Q_{i_{max}}(n)\right\}.
\eeq
Hence, we obtain
\begin{multline}
\bm{\lambda}\cdot\mf{Q}(n) -\sum_{\{\mf{a}\in\mc{A},\mf{a}\neq \mf{0}\}}\mb{P}(\mf{T}(n)=\mf{a})\max_{\bm{\pi}\in \mc{M}}\left\{\mf{r}(\mf{a},\bm{\pi})\cdot \mf{Q}(n)\right\}\\
\leq
-\delta\sum_{\{\mf{a}\in\mc{A},\mf{a}\neq \mf{0}\}}\mb{P}(\mf{T}(n)=\mf{a})\max_{\bm{\pi}\in \mc{M}}\left\{r_{i_{max}}(\mf{a},\bm{\pi})Q_{i_{max}}(n)\right\}.\nonumber
\end{multline}
We can write
\begin{multline}
\bm{\lambda}\cdot\mf{Q}(n) -\sum_{\{\mf{a}\in\mc{A},\mf{a}\neq \mf{0}\}}\mb{P}(\mf{T}(n)=\mf{a})\max_{\bm{\pi}\in \mc{M}}\left\{\mf{r}(\mf{a},\bm{\pi})\cdot \mf{Q}(n)\right\}\\
\leq
-\delta\max_{\{\mf{a}\in\mc{A},\mf{a}\neq \mf{0}\}}\mb{P}(\mf{T}(n)=\mf{a})\left\{
\max_{\bm{\pi}\in \mc{M}}r_{i_{max}}(\mf{a},\bm{\pi})Q_{i_{max}}(n)\right\}.\nonumber
\end{multline}
Now we write
\begin{multline}
\bm{\lambda}\cdot\mf{Q}(n) -\sum_{\{\mf{a}\in\mc{A},\mf{a}\neq \mf{0}\}}\mb{P}(\mf{T}(n)=\mf{a})\max_{\bm{\pi}\in \mc{M}}\left\{\mf{r}(\mf{a},\bm{\pi})\cdot \mf{Q}(n)\right\}\\
\leq
-\delta[\min_i\max_{\{\mf{a}\in\mc{A},\mf{a}\neq \mf{0}\}}\mb{P}(\mf{T}(n)=\mf{a})\max_{\bm{\pi}\in \mc{M}}r_{i}(\mf{a},\bm{\pi})]Q_{i_{max}}(n).\nonumber
\end{multline}
Finally, using the fact that $Q_{i_{max}}(n)\geq \frac{\norm{\mf{Q}(n)}}{\sqrt{M}}$, we obtain
\beq
\bm{\lambda}\cdot\mf{Q}(n) -\sum_{\{\mf{a}\in\mc{A},\mf{a}\neq \mf{0}\}}\mb{P}(\mf{T}(n)=\mf{a})\max_{\bm{\pi}\in \mc{M}}\left\{\mf{r}(\mf{a},\bm{\pi})\cdot \mf{Q}(n)\right\}
\leq
-\delta \beta\frac{\norm{\mf{Q}(n)}}{\sqrt{M}},\nonumber
\eeq
where 
\beq
\beta=\min_i\max_{\{\mf{a}\in\mc{A},\mf{a}\neq \mf{0}\}}\mb{P}(\mf{T}(n)=\mf{a})\max_{\bm{\pi}\in \mc{M}}\indic{\bm{\pi}\in \mc{S}(\mf{a}}u_{i}(\mf{a},\bm{\pi})\nonumber.
\eeq
From \eqref{eq:drift_merge}, we have
\beq
    \expect{V(\mf{Q}(n+1))-V(\mf{Q}(n))\given \mf{Q}(n)}
    \leq \sum_{i=1}^M\expect{A_i(n)^2}+M
-2\delta \beta\frac{\norm{\mf{Q}(n)}}{\sqrt{M}}.\nonumber
    \eeq
    Let $B$ be defined as
    \beq B=\sum_{i=1}^M\expect{A_i(n)^2}+M\nonumber.
\eeq
For $\epsilon=\frac{\delta \beta}{\sqrt{M}}$,
if $\norm{\mf{Q}(n)}\geq \frac{2B}{\epsilon}$, then we obtain
\beq
    \expect{V(\mf{Q}(n+1))-V(\mf{Q}(n))\given \mf{Q}(n)}
    \leq 
-\frac{3\epsilon}{2}\norm{\mf{Q}(n)}.\nonumber
    \eeq
    Using the Lyapunov stability theorem \cite[Theorem~3.1]{Leandros_maxweight}, we conclude that the switch is stable.
    This completes the proof.

\section{Conclusions}
\label{sec:conclusions}
In this paper, we have investigated stability properties of a quantum switch that provide insights into performance of the switch. We proposed a Max-Weight scheduling policy that takes into accounts for differences in various parameters so as to achieve good performance. We also proved that the proposed policy stabilizes the switch for all feasible arrival rates.
Although our policy has high implementation cost due to the fact that it requires the switch to search over all possible matchings to find the best matching in each time-slot, it provides insights into how to design low complexity scheduling algorithms and also its performance acts as a benchmark to the performance of other policies.

We plan to address several important problems in future work.
We would like to investigate the design and analysis of scheduling algorithms that have low implementation costs. We also plan to study the case where Bell-pairs take more than one time-slot to decohere.
Finally, it is of interest to analyze scheduling algorithms for distribution of entangled states over quantum networks and also consider the effect of entanglement purification procedures into the design of scheduling algorithms.

\section*{ACKNOWLEDGMENTS}     
This work was supported in part by the National Science Foundation under Grants CNS-1955834 and ERC-1941583.
 

\bibliographystyle{IEEEtran}
\bibliography{IEEEabrv,reference_spie21} 

\begin{thebibliography}{10}
\providecommand{\url}[1]{#1}
\csname url@samestyle\endcsname
\providecommand{\newblock}{\relax}
\providecommand{\bibinfo}[2]{#2}
\providecommand{\BIBentrySTDinterwordspacing}{\spaceskip=0pt\relax}
\providecommand{\BIBentryALTinterwordstretchfactor}{4}
\providecommand{\BIBentryALTinterwordspacing}{\spaceskip=\fontdimen2\font plus
\BIBentryALTinterwordstretchfactor\fontdimen3\font minus
  \fontdimen4\font\relax}
\providecommand{\BIBforeignlanguage}[2]{{%
\expandafter\ifx\csname l@#1\endcsname\relax
\typeout{** WARNING: IEEEtran.bst: No hyphenation pattern has been}%
\typeout{** loaded for the language `#1'. Using the pattern for}%
\typeout{** the default language instead.}%
\else
\language=\csname l@#1\endcsname
\fi
#2}}
\providecommand{\BIBdecl}{\relax}
\BIBdecl

\bibitem{Bennett2014}
C.~H. Bennett and G.~Brassard, ``Quantum cryptography: Public key distribution
  and coin tossing,'' \emph{Theor. Comput. Sci.}, vol. 560, pp. 7--11, 2014.

\bibitem{Ekert}
A.~K. Ekert, ``Quantum cryptography based on bell's theorem,'' \emph{Phys. Rev.
  Lett.}, vol.~67, pp. 661--663, Aug 1991.

\bibitem{Eldredge}
Z.~Eldredge, M.~Foss-Feig, J.~A. Gross, S.~L. Rolston, and A.~V. Gorshkov,
  ``Optimal and secure measurement protocols for quantum sensor networks,''
  \emph{Phys. Rev. A}, vol.~97, p. 042337, Apr 2018.

\bibitem{Giovannetti_2011}
\BIBentryALTinterwordspacing
V.~Giovannetti, S.~Lloyd, and L.~Maccone, ``Advances in quantum metrology,''
  \emph{Nature Photonics}, vol.~5, no.~4, pp. 222--229, mar 2011. [Online].
  Available: \url{https://doi.org/10.1038%2Fnphoton.2011.35}
\BIBentrySTDinterwordspacing

\bibitem{Xia:2019jil}
Y.~Xia, W.~Li, W.~Clark, D.~Hart, Q.~Zhuang, and Z.~Zhang, ``{Demonstration of
  a Reconfigurable Entangled Radiofrequency-Photonic Sensor Network},''
  \emph{Phys. Rev. Lett.}, vol. 124, no.~15, p. 150502, 2020.

\bibitem{Broadbent}
A.~Broadbent, J.~Fitzsimons, and E.~Kashefi,
  ``\BIBforeignlanguage{English}{Universal blind quantum computation},'' in
  \emph{\BIBforeignlanguage{English}{Proceedings of the 50th Annual IEEE
  Symposium on Foundations of Computer Science (FOCS '09)}}, ser. Annual
  Symposium on Foundations of Computer Science.\hskip 1em plus 0.5em minus
  0.4em\relax United States: Institute of Electrical and Electronics Engineers
  (IEEE), 2009, pp. 517--526.

\bibitem{Pirandola_2019}
\BIBentryALTinterwordspacing
S.~Pirandola, ``End-to-end capacities of a quantum communication network,''
  \emph{Communications Physics}, vol.~2, no.~1, May 2019. [Online]. Available:
  \url{http://dx.doi.org/10.1038/s42005-019-0147-3}
\BIBentrySTDinterwordspacing

\bibitem{Pant}
M.~Pant, H.~Krovi, D.~Towsley, L.~Tassiulas, L.~Jiang, P.~Basu, D.~Englund, and
  S.~Guha, ``\BIBforeignlanguage{English (US)}{Routing entanglement in the
  quantum internet},'' \emph{\BIBforeignlanguage{English (US)}{npj Quantum
  Information}}, vol.~5, no.~1, Dec. 2019.

\bibitem{Dahlberg_2019}
\BIBentryALTinterwordspacing
A.~Dahlberg, M.~Skrzypczyk, T.~Coopmans, L.~Wubben, F.~Rozpundefineddek,
  M.~Pompili, A.~Stolk, P.~Pawe\l{}czak, R.~Knegjens, J.~de~Oliveira~Filho,
  R.~Hanson, and S.~Wehner, ``A link layer protocol for quantum networks,'' in
  \emph{Proceedings of the ACM Special Interest Group on Data Communication},
  ser. SIGCOMM '19.\hskip 1em plus 0.5em minus 0.4em\relax New York, NY, USA:
  Association for Computing Machinery, 2019, p. 159-173. [Online]. Available:
  \url{https://doi.org/10.1145/3341302.3342070}
\BIBentrySTDinterwordspacing

\bibitem{Van_meter}
R.~{Van Meter}, \emph{\BIBforeignlanguage{English}{Quantum Networking}}.\hskip
  1em plus 0.5em minus 0.4em\relax Wiley Blackwell, Jun. 2014, vol.
  9781848215375.

\bibitem{Bhaskar_2020}
M.~K. Bhaskar, R.~Riedinger, B.~Machielse, D.~S. Levonian, C.~T. Nguyen, E.~N.
  Knall, H.~Park, D.~R. Englund, M.~Lon{\v{c}}ar, D.~D. Sukachev, and M.~D.
  Lukin, ``Experimental demonstration of memory-enhanced quantum
  communication,'' \emph{Nature}, vol. 580, pp. 60--64, 2020.

\bibitem{lee2020quantum}
Y.~Lee, E.~Bersin, A.~Dahlberg, S.~Wehner, and D.~Englund, ``A quantum router
  architecture for high-fidelity entanglement flows in quantum networks,''
  2020.

\bibitem{Ruoyu}
R.~Li, L.~Petit, D.~Franke, J.~Dehollain, J.~Helsen, M.~Steudtner, N.~Thomas,
  S.~Wehner, L.~Vandersypen, and M.~Veldhorst, ``\BIBforeignlanguage{English}{A
  crossbar network for silicon quantum dot qubits},''
  \emph{\BIBforeignlanguage{English}{Science Advances}}, vol.~4, no.~7, Jul.
  2018.

\bibitem{Armstrong_2012}
\BIBentryALTinterwordspacing
S.~Armstrong, J.-F. Morizur, J.~Janousek, B.~Hage, N.~Treps, P.~K. Lam, and
  H.-A. Bachor, ``Programmable multimode quantum networks,'' \emph{Nature
  Communications}, vol.~3, no.~1, Jan 2012. [Online]. Available:
  \url{http://dx.doi.org/10.1038/ncomms2033}
\BIBentrySTDinterwordspacing

\bibitem{Herbauts:13}
\BIBentryALTinterwordspacing
I.~Herbauts, B.~Blauensteiner, A.~Poppe, T.~Jennewein, and H.~H\"{u}bel,
  ``Demonstration of active routing of entanglement in a multi-user network,''
  \emph{Opt. Express}, vol.~21, no.~23, pp. 29\,013--29\,024, Nov 2013.
  [Online]. Available:
  \url{http://www.opticsexpress.org/abstract.cfm?URI=oe-21-23-29013}
\BIBentrySTDinterwordspacing

\bibitem{Hall}
\BIBentryALTinterwordspacing
M.~A. Hall, J.~B. Altepeter, and P.~Kumar, ``Ultrafast switching of photonic
  entanglement,'' \emph{Phys. Rev. Lett.}, vol. 106, p. 053901, Feb 2011.
  [Online]. Available:
  \url{https://link.aps.org/doi/10.1103/PhysRevLett.106.053901}
\BIBentrySTDinterwordspacing

\bibitem{nielsen00}
M.~A. Nielsen and I.~L. Chuang, \emph{Quantum Computation and Quantum
  Information}.\hskip 1em plus 0.5em minus 0.4em\relax Cambridge University
  Press, 2000.

\bibitem{Boxi_li}
B.~Li, T.~Coopmans, and D.~Elkouss, ``Efficient optimization of cut-offs in
  quantum repeater chains,'' ser. Proceedings - IEEE International Conference
  on Quantum Computing and Engineering, QCE 2020, 2020, pp. 158--168.

\bibitem{Burmeister:08}
\BIBentryALTinterwordspacing
E.~F. Burmeister, J.~P. Mack, H.~N. Poulsen, J.~Klamkin, L.~A. Coldren, D.~J.
  Blumenthal, and J.~E. Bowers, ``Soa gate array recirculating buffer with
  fiber delay loop,'' \emph{Opt. Express}, vol.~16, no.~12, pp. 8451--8456, Jun
  2008. [Online]. Available:
  \url{http://www.opticsexpress.org/abstract.cfm?URI=oe-16-12-8451}
\BIBentrySTDinterwordspacing

\bibitem{Shchukin}
\BIBentryALTinterwordspacing
E.~Shchukin, F.~Schmidt, and P.~van Loock, ``Waiting time in quantum repeaters
  with probabilistic entanglement swapping,'' \emph{Phys. Rev. A}, vol. 100, p.
  032322, Sep 2019. [Online]. Available:
  \url{https://link.aps.org/doi/10.1103/PhysRevA.100.032322}
\BIBentrySTDinterwordspacing

\bibitem{Gayane1}
\BIBentryALTinterwordspacing
G.~Vardoyan, S.~Guha, P.~Nain, and D.~Towsley, ``{On the Capacity Region of
  Bipartite and Tripartite Entanglement Switching},'' in \emph{{Performance
  2020 - 38th IFIP International Symposium on Computer Performance, Modeling,
  Measurements and Evaluation}}, Milan, Italy, Nov. 2020, pp. 1--6. [Online].
  Available: \url{https://hal.inria.fr/hal-02010865}
\BIBentrySTDinterwordspacing

\bibitem{Gayane2}
\BIBentryALTinterwordspacing
P.~Nain, G.~Vardoyan, S.~Guha, and D.~Towsley, ``On the analysis of a
  multipartite entanglement distribution switch,'' \emph{Proc. ACM Meas. Anal.
  Comput. Syst.}, vol.~4, no.~2, Jun. 2020. [Online]. Available:
  \url{https://doi.org/10.1145/3392141}
\BIBentrySTDinterwordspacing

\bibitem{Guha_repeater}
\BIBentryALTinterwordspacing
S.~Guha, H.~Krovi, C.~A. Fuchs, Z.~Dutton, J.~A. Slater, C.~Simon, and
  W.~Tittel, ``Rate-loss analysis of an efficient quantum repeater
  architecture,'' \emph{Phys. Rev. A}, vol.~92, p. 022357, Aug 2015. [Online].
  Available: \url{https://link.aps.org/doi/10.1103/PhysRevA.92.022357}
\BIBentrySTDinterwordspacing

\bibitem{dhara2021subexponential}
P.~Dhara, A.~Patil, H.~Krovi, and S.~Guha, ``Sub-exponential rate versus
  distance with time multiplexed quantum repeaters,'' 2021.

\bibitem{dhara2021multiplexed}
P.~Dhara, N.~M. Linke, E.~Waks, S.~Guha, and K.~P. Seshadreesan, ``Multiplexed
  quantum repeaters based on dual-species trapped-ion systems,'' 2021.

\bibitem{Leandros_maxweight}
L.~Tassiulas and A.~Ephremides, ``Stability properties of constrained queueing
  systems and scheduling policies for maximum throughput in multihop radio
  networks,'' in \emph{29th IEEE Conference on Decision and Control}, 1990, pp.
  2130--2132 vol.4.

\bibitem{McKeown}
N.~McKeown, A.~Mekkittikul, V.~Anantharam, and J.~Walrand, ``Achieving 100\%
  throughput in an input-queued switch,'' \emph{IEEE Transactions on
  Communications}, vol.~47, no.~8, pp. 1260--1267, 1999.

\bibitem{srikant_book}
\BIBentryALTinterwordspacing
R.~Srikant and L.~Ying, \emph{Communication Networks: An Optimization, Control
  and Stochastic Networks Perspective}.\hskip 1em plus 0.5em minus 0.4em\relax
  Cambridge University Press, 2014. [Online]. Available:
  \url{https://books.google.co.uk/books?id=Aa\_CAQAAQBAJ}
\BIBentrySTDinterwordspacing

\bibitem{Leandros_Varying}
L.~Tassiulas, ``Scheduling and performance limits of networks with constantly
  changing topology,'' \emph{IEEE Transactions on Information Theory}, vol.~43,
  no.~3, pp. 1067--1073, 1997.

\bibitem{Kavita_maxweight}
M.~Andrews, K.~Kumaran, K.~Ramanan, A.~Stolyar, R.~Vijayakumar, and P.~Whiting,
  ``Scheduling in a queuing system with asynchronously varying service rates,''
  \emph{Probability in the Engineering and Informational Sciences}, vol.~18,
  no.~2, pp. 191--217, 2004.

\end{thebibliography}

\end{document}